# A Surface-Based Federated Chow Test Model for Integrating APOE Status, Tau Deposition Measure, and Hippocampal Surface Morphometry

Running title: Surface-based federated Chow test Model

Jianfeng Wu[a], Yi Su[b], Yanxi Chen[a], Wenhui Zhu[a], Eric M. Reiman[b], Richard J. Caselli[d], Kewei Chen[b], Paul M. Thompson[e], Junwen Wang[c], Yalin Wang[a]
and for the Alzheimer's Disease Neuroimaging Initiative*

[a] School of Computing and Augmented Intelligence, Arizona State University, Tempe, USA;
[b] Banner Alzheimer's Institute, Phoenix, USA;
[c] Division of Applied Oral Sciences & Community Dental Care, Faculty of Dentistry, The University of Hong Kong, Hong Kong SAR, China
[d] Department of Neurology, Mayo Clinic Arizona, Scottsdale, USA;
[e] Imaging Genetics Center, Stevens Neuroimaging and Informatics Institute, University of Southern California, Marina del Rey, USA

Please address correspondence to:

Dr. Yalin Wang
School of Computing and Augmented Intelligence
Arizona State University
P.O. Box 878809
Tempe, AZ 85287 USA
Phone: (480) 965-6871
Fax: (480) 965-2751
E-mail: ylwang@asu.edu

*Acknowledgments: Data used in preparing this article were obtained from the Alzheimer's Disease Neuroimaging Initiative (ADNI) database (adni.loni.usc.edu). As such, many investigators within the ADNI contributed to the design and implementation of ADNI and/or provided data but did not participate in the analysis or writing of this report. A complete listing of ADNI investigators can be found at: http://adni.loni.usc.edu/wp-content/uploads/how_to_apply/ADNI_Acknowledgement_List.pdf.

# ABSTRACT

**Background:** Alzheimer's disease (AD) is the most common type of age-related dementia, affecting 6.2 million people aged 65 or older according to CDC data. It is commonly agreed that discovering an effective AD diagnosis biomarker could have enormous public health benefits, potentially preventing or delaying up to 40% of dementia cases. Tau neurofibrillary tangles are the primary driver of downstream neurodegeneration and subsequent cognitive impairment in AD, resulting in structural deformations such as hippocampal atrophy that can be observed in magnetic resonance imaging (MRI) scans.

**Objective:** To build a surface-based model to 1) detect differences between APOE subgroups in patterns of tau deposition and hippocampal atrophy, and 2) use the extracted surface-based features to predict cognitive decline.

**Methods:** Using data obtained from different institutions, we develop a surface-based federated Chow test model to study the synergistic effects of APOE, a previously reported significant risk factor of AD, and tau on hippocampal surface morphometry.

**Results:** We illustrate that the APOE-specific morphometry features correlate with AD progression and better predict future AD conversion than other MRI biomarkers. For example, a strong association between atrophy and abnormal tau was identified in hippocampal subregion cornu ammonis 1 (CA1 subfield) and subiculum in e4 homozygote cohort.

**Conclusion:** Our model allows for identifying MRI biomarkers for AD and cognitive decline prediction and may uncover a corner of the neural mechanism of the influence of APOE and tau deposition on hippocampal morphology.

## 1. INTRODUCTION

Alzheimer's disease (AD) is one of the most common causes of dementia in the elder population, whose process begins years before clinical symptoms. Therefore, improving the diagnostic accuracy of AD is helpful in the early detection of AD and may provide remarkable social benefits. Researchers have been using brain biomarkers to facilitate the accuracy and efficiency of AD detection [1], with Amyloid-β (Aβ) plaques and tau tangles being the two most popular pathological hallmarks of AD. Specifically, those markers are not only crucial in dementia pathogenesis but are also significantly related to structural deformations, which can be physically determined by magnetic resonance imaging (MRI) scans [2]. In the A/T/N framework, the presence of AD pathology is defined by the abnormal levels of Aβ in the brain or cerebrospinal fluid (CSF), and AD is defined by additional abnormal tau levels without considering symptomatology [3]. Specifically, the tau protein tangles accumulate where Aβ is produced faster than eliminated, potentially resulting in atrophy and neurodegeneration that is detectable by MRI [4,5]. One of the primary AD pathology targets from clinically normal status to dementia is the hippocampus, which has also been identified as the target in the earliest disease stages [6,7].

Apolipoprotein E (*APOE*) genotype has been identified as one of the strongest factors influencing AD risk. In particular, the e4 and e2 alleles are associated with increased and decreased risk, respectively. In the Caucasian population, individuals carrying two e4 alleles have almost five folds increased risk of developing AD (odds ratio increases from 3.2 to 14.9) but carrying one e2 allele decreases the OR to only 0.6 [8].

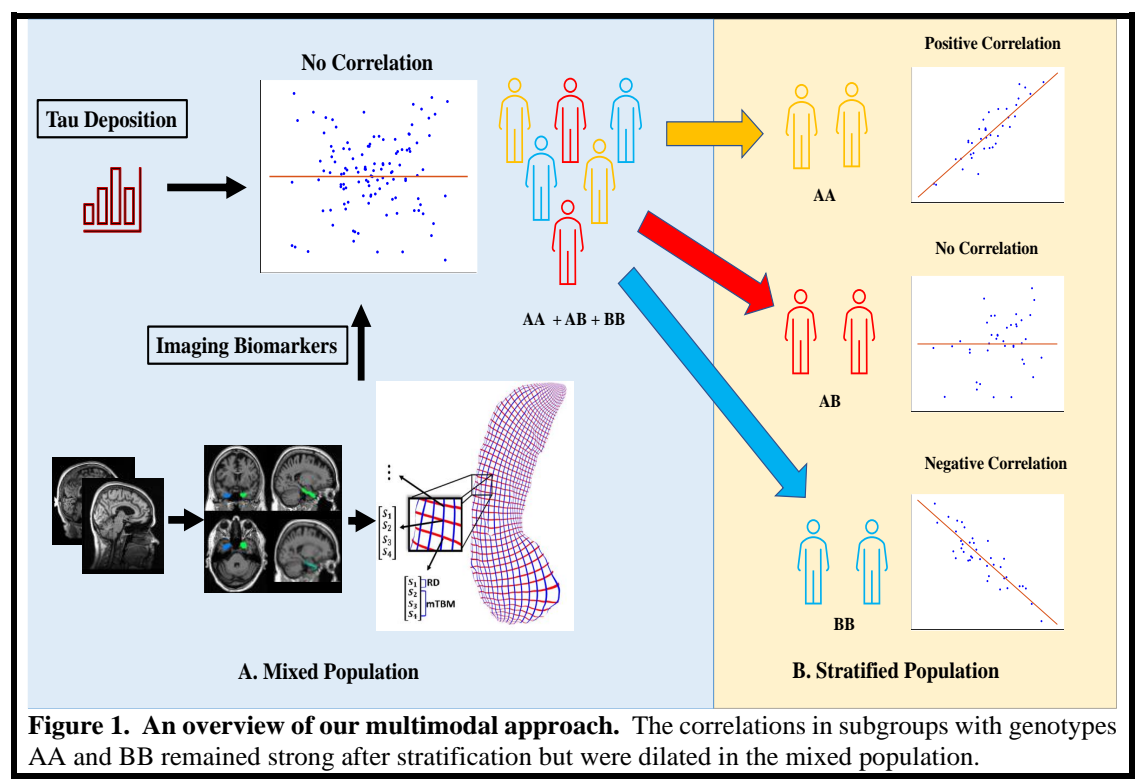

**Figure 1. An overview of our multimodal approach.** The correlations in subgroups with genotypes AA and BB remained strong after stratification but were dilated in the mixed population.

Progressive Aβ and tau accumulation are more prominent in e4 carriers [7]. It is less clear, however, whether patterns of tau-induced morphological changes in the hippocampus differ by APOE genotype and whether any such differences might prove to be a more accurate predictor than the overall hippocampal volume of disease progression, regardless of the high variability among individual patients. Current AD clinical research utilizes imaging biomarker outcomes that often include not only structural MRI but also tau PET. Clinical research data are often gathered from multiple sites and are limited in size. Yet, different institutions can hardly share research data due to patient privacy concerns or legal complexities, such as data restrictions based on patient consent or institutional review board (IRB) regulations. Nonetheless, successfully integrating data from multiple sites and sources can achieve larger sample sizes that increase statistical power for detecting true therapeutic effects. Recently, federated learning approaches [9–11] have been actively studied to address this distributed problem. However, most meta-analytic neuroimaging studies, e.g., the ENIGMA consortium [12], currently focus on univariate measures derived from brain MRI, diffusion tensor imaging, electroencephalogram, or other data modalities, but few have integrated genetics, MRI, and tau PET data to infer the multimodal relationships we hypothesize to exist.

Our study has both a scientific and a methodological goal. Scientifically, we will test the hypothesis that hippocampal morphological alterations will differ by APOE genotype, that in turn may reveal more sensitive structural outcomes that better correlate with disease progression and, therefore treatment effects. Methodologically, we pursue an effective multimodal data fusion strategy, a surface-based federated model based on the Chow test [13,14] that allows us to integrate incomplete data sets from multiple sources, resulting in greater statistical power to address our scientific hypothesis. The intuition of our multimodal approach is illustrated in **Figure 1**. The image-tau relationship (correlation) is diluted in a genotypically mixed population, but strong correlations (AA and BB groups) across subgroups are still observable after stratification on genotypes. Patient data from the Alzheimer Disease

Neuroimaging Initiative (ADNI) cohort [14] were first stratified into three sub-cohorts according to their *APOE* genotypes, including non-carriers (NC), heterozygote subjects (HT), and homozygote subjects (HM). We then randomly assigned the subjects to five hypothetical institutions. Then, we used our surface-based federated Chow test model to evaluate the correlations between the measure for tau deposition/regional uptakes of different Braak ROIs and imaging biomarkers. In this study, we consider Braak 1/34/56 ROIs. Braak 2 was excluded because of a previously reported 18F-Flortaucipir off-target binding in this region [15,16]. In the Chow test model, each imaging biomarker was used as the predictor, and the measure for tau was used as the response, as shown in **Figure 2**.

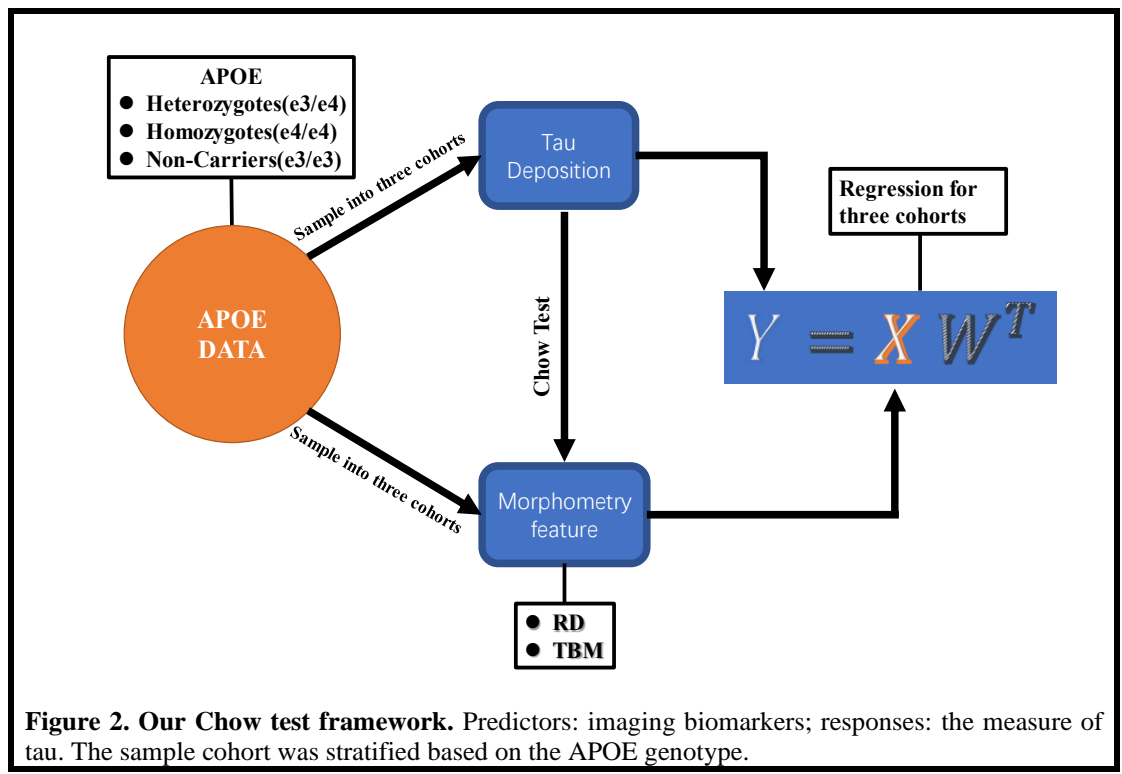

**Figure 2. Our Chow test framework.** Predictors: imaging biomarkers; responses: the measure of tau. The sample cohort was stratified based on the APOE genotype.

We first fit a Chow test model with hippocampal volume as a predictor, commonly used by other studies. We further applied two morphometry features as the imaging biomarker, radial distance (RD) and surface tensor-based morphometry (TBM), to determine the regions of interest (ROIs), the surface areas with the greatest degree of atrophy. In addition, we evaluated the top features of the selected ROIs with survival analysis on a separate dataset. We hypothesize that our model will detect the difference in the association between tau deposition and hippocampal atrophy among the cohorts with different genotypes and, therefore, extract the most AD-related regions on the hippocampal surfaces.

## 2. SUBJECTS and METHODS

### 2.1 Subjects

Data used in the preparation of this article were obtained from the Alzheimer's Disease Neuroimaging Initiative (ADNI) database (adni.loni.usc.edu, [14]). The ADNI was launched in 2003 and has gone through multiple phases of development, including ADNI, ADNI 1, ADNI 2, ADNI GO and ADNI 3. Nowadays, ADNI has become an integrated data repository providing measurements in biological markers, clinical and neuropsychological assessments, which can be downloaded and used either separately or combined for AD-related research. For more detailed information, see www.adni-info.org. In this study, we obtained 847 pairs of MRI scans and AV1451 PET images, including 502 non-carriers (NC), 281 heterozygote subjects (HT) and 64 homozygote subjects (HM). **Table 1** shows the demographic information from the cohort that we analyzed.

For flortaucipir tau-PET, tau data are reprocessed using a single pipeline consistent with [17] so that the standardized uptake value ratio (SUVR) from different ADNI study sites can be analyzed together. In this work, we examine the regional SUVR for tau deposition, corresponding to Braak stages 1/34/56 ROIs [18–21].

Table 1. Demographic information for the subjects we study from the ADNI.

| Group | Diagnosis (NL/MCI/AD) | Age | Sex (M/F) | Braak1 | Braak34 | Braak56 | MMSE |
|---|---|---|---|---|---|---|---|
| NC (N=502) | 320/139/43 | 75.4(±7.5) | 242/260 | 2.17(±0.76) | 1.75(±0.33) | 1.80(±0.33) | 28.2(±2.8) |
| HT (N=281) | 175/71/35 | 73.4(±7.2) | 121/160 | 2.39(±0.86) | 1.89(±0.51) | 1.88(±0.41) | 27.8(±3.1) |
| HM (N=64) | 12/30/22 | 71.0(±7.9) | 33/31 | 2.86(±0.76) | 2.18(±0.68) | 2.23(±0.80) | 25.8(±4.6) |

Values are mean ± standard deviation where applicable.

### 2.2 Proposed Pipeline

In this work, we propose a Federated Chow test model to study the synergistic effects of APOE and tau on hippocampal morphometry, as illustrated in **Figure 3**. **Figure 3** (a-d) shows the image preprocessing step finished at each local institution. Each institution first extracts the morphometric features from MRI scans locally. After image registration of the available MR images, the hippocampal structures are segmented and smoothed for surface generation. After the surface parameterization and fluid registration, the hippocampal radial distance (RD) and tensor-based morphometry (TBM) features are calculated at each surface point. Then, the federated Chow test model is performed on each morphometry feature across these institutions, and a *p*-value is acquired for each feature on each vertex. After calculating all the *p*-values, we further visualized them on the hippocampal surface.

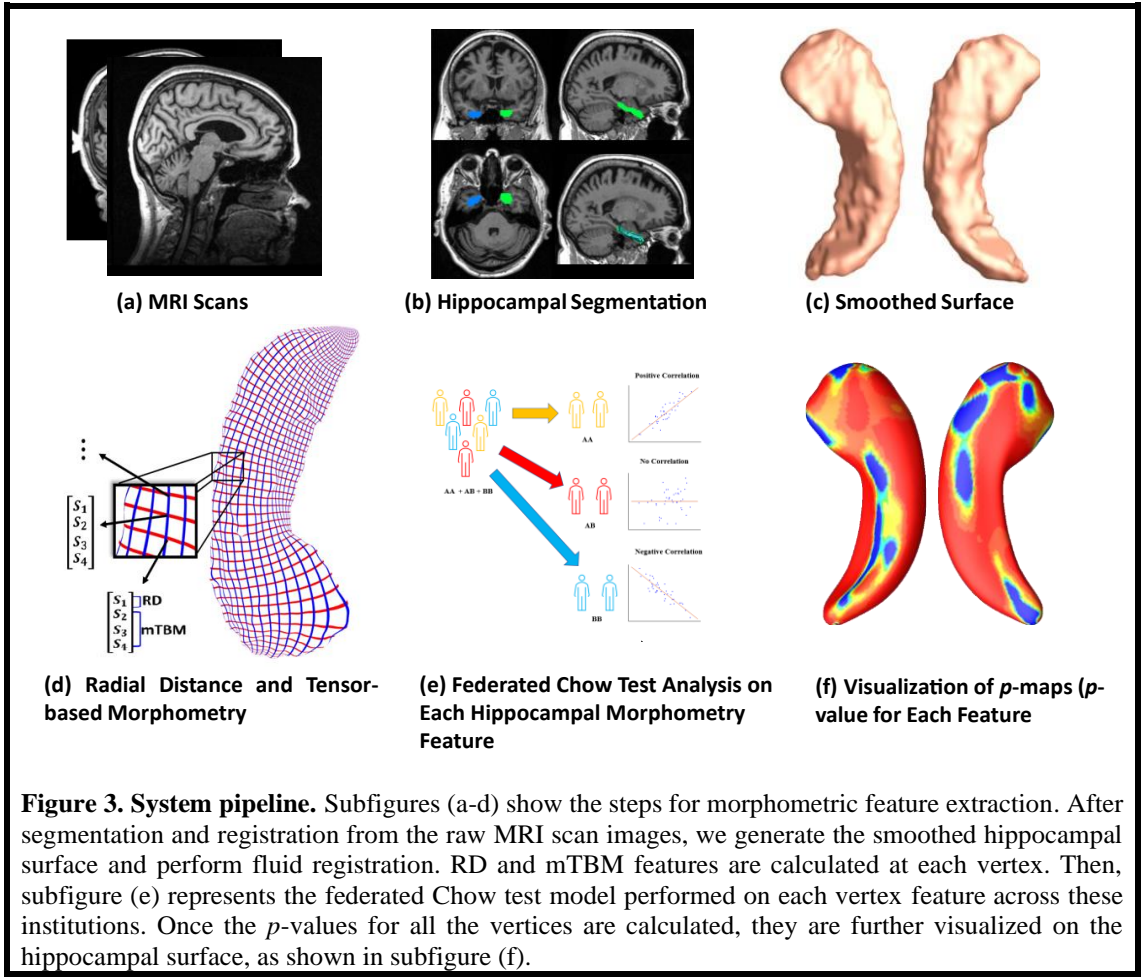

**Figure 3. System pipeline.** Subfigures (a-d) show the steps for morphometric feature extraction. After segmentation and registration from the raw MRI scan images, we generate the smoothed hippocampal surface and perform fluid registration. RD and mTBM features are calculated at each vertex. Then, subfigure (e) represents the federated Chow test model performed on each vertex feature across these institutions. Once the *p*-values for all the vertices are calculated, they are further visualized on the hippocampal surface, as shown in subfigure (f).

### 2.2.1 Image Preprocessing

For hippocampus segmentation and hippocampal surface reconstruction, we adopt the protocol from our prior work [22]. Specifically, we use FIRST (FMRIB's Integrated Registration and Segmentation Tool) [23] to segment the hippocampus substructure (Figure 3 (b)). We reconstruct hippocampal surfaces with a topology-preserving level set method [24] and the marching cubes algorithm [25], followed by our surface smoothing process consists of mesh simplification using progressive meshes [26] and mesh refinement by the Loop subdivision surface method [27] (Figure 3 (c)). The smoothed meshes are accurate approximations of the original surfaces with a higher signal-to-noise ratio (SNR).

We compute the conformal grid (150×100) on each hippocampus surface using a holomorphic 1-form basis [28,29] and use surface conformal representation [30,31] and registered the hippocampus surface to a template surface with our surface fluid registration method [31].

A one-to-one correspondence map between hippocampal surfaces was established after parameterization and registration, where each surface has the same number of vertices (150 × 100). As in our previous methods [32], we adopted two morphometry features: radial distance (RD) [33,34] and tensor-based morphometry (TBM) [34]. The RD (a scalar at each vertex) primarily reflects surface differences along the surface normal directions. The medial axis is determined by the geometric center of the isoparametric curve on the computed conformal grid [32]. The axis is perpendicular to the isoparametric curve, so the thickness can be easily calculated as the Euclidean distance between the core and the vertex on the curve. TBM examines the Jacobian matrix $J$ of the deformation map that registers the surface to a template surface [31]. As a result, with 15,000 vertices in both left and right hippocampus, a total of 60,000 = (15,000 + 15,000) * 2 features are generated.

### 2.2.2 Federated Chow Test Analysis

In 1960, econometrician Gregory Chow proposed Chow test, which determines whether the true coefficients in two regressions are equal [12]. In situations where the whole group of samples are split into subgroups, the Chow test evaluates the equality of the correlation coefficient across the subgroups. For example, a typical linear regression model appears in the form of $y = wX + \epsilon$. Breaking the data into two subgroups results in two separate regressions:

$y_1 = w_1 x_1 + \epsilon$ and $y_2 = w_2 x_2 + \epsilon$. Then, the Chow test statistic is an F-statistic with $k$ and $N_1 + N_2 - 2k$ degrees of freedom: $F = \frac{(S_C - (S_1 + S_2))/k}{(S_1 + S_2)/(N_1 + N_2 - 2k)}$, where k is the number of parameters, $N_1$ and $N_2$ are the sample number in each subgroup, and $S_C, S_1, S_2$ are the sum of squared residuals from the combined regression and two separate regressions, respectively.

Based on the Chow test model, our recent work [35] proposed federated Chow test model, Genotype-Expression-Imaging Data Integration (GEIDI), to identify genetic and transcriptomic influences on brain sMRI measures. In this work, we further generalize it to surface models to leverage the surface model's visualization and geometric sensitivity. We apply the proposed model to the ADNI dataset to detect the synergistic effects of APOE and tau on hippocampal morphometry.

We simulate the multi-site condition by separating all the samples into $I$ hypothetical institutions ($I = 5$) on a cluster with several conventional x86 nodes, of which each contains two Intel Xeon E5-2680 v4 CPUs running at 2.40 GHz. Each parallel computing node has a full-speed Omni-Path connection to every other node in its partition. (Although the ADNI data can be centralized, such a federated analysis would allow the method to be scaled up to much larger datasets, including genomic data that is difficult to centralize for logistic or regulatory reasons). Each institution can further stratify the local subjects into subgroups based on their genotypes. The image biomarkers and gene expression values can be represented as $X_i^g$ and $y_i^g$ respectively, where $i$ is the institution index and $g$ is the subgroup index within $i$th institution (**Figure 2**). All samples from the same group across all institutions will be fit into a regression model.

Using federated linear regression, we can calculate four linear models for all the $I$ institutions, including three models for three subgroups and one for all samples in the three subgroups. For details of federated linear model, see [35].

### 2.2.3 Surface-based Federated Linear Regression

Since hippocampal surfaces are registered in our work, we can apply the Federated Chow test model on each surface point. It is a natural generalization of our prior federated Chow test [35] to a surface-based federated Chow test model.

To avoid overfitting and the requirement of additional regularization parameters, we choose linear regression [36] as our base regression model in Chow test, although other models such as polynomial regression [37] or ridge regression [38] are also potential alternatives. In each subgroup, a linear regression is calculated as: $y = Xw + \epsilon$, where $X \in R^{N \times k}$ represents the independent variables, $y \in R^N$ is a vector of the observations on a dependent variable, $w \in R^k$ is a coefficient vector, and $\epsilon \in R^N$ is the disturbance vector. $N$ is the number of observations in the group, and $k$ is the number of parameters. In our experiments, $k$ is equal to 1 since we only evaluate one feature (one RD or one TBM) on each vertex. Then, the coefficient vector $w$ can be estimated by minimizing the least squared function, $S(w) = \frac{1}{2}\|Xw - y\|_2^2$.

To avoid centralizing the data, $(X_i, y_i)$, from each institution, we first rewrite the minimization problem as, $\min_{w} \sum_{i=1}^{I} S_i(w; X_i, y_i) = \frac{1}{2}\sum_{i=1}^{I}\|X_i w - y_i\|_2^2$. Then, the global gradient can be calculated as, $\nabla S(w) = X^T(Xw - y) = \sum_{i=1}^{I} X_i^T(X_i w - y_i) = \sum_{i=1}^{I} \nabla S_i(w)$. In this way, each institution can compute $\nabla S_i(w)$ locally before sending the result to the global center, and the global center sends the result back after gathering all the partial gradients and computing the global gradient $\nabla S(w)$. Finally, local institutions update $w$ by: $w \leftarrow w - \eta \nabla S(w)$. The whole framework of our federated Genotype-Expression-Image Integration model is summarized in the Appendix as **Algorithm 1**.

### 2.3 Performance Evaluation Protocol

The PASCS-MP pipeline contains a set of parameters including the number of patches, patch size, dimensionality of sparse coding, l1-norm regularization parameters and the kernel size. Our previous work has determined the best combination of those parameters on hippocampus surfaces, which were adopted in this study as the default values [38]. We instead focus on the performance evaluation of our models with or without extracted ROI features. Specifically, we compared the prediction accuracy, sensitivity, specificity and AUC of each model using different available features, including area, volume, RD/TBM of whole hippocampus, and RD/TBM of ROIs. In addition, we fit a COX proportional hazard model and calculated the p-values of each model, where lower p-values indicate higher significance of the model.

## 3. RESULTS

### 3.1 Linking Hippocampal Volume to Tau Deposition

To fit the Chow test model [35] to evaluate the difference between hippocampal volume change and the tau measure of different Braak ROIs, we first stratified the total of 847 subjects into three subgroups based on their APOE genotypes, including non-carriers (NC), heterozygotes (HT) and homozygotes (HM). The federated Chow test model [35] was then used to explore the difference between the changes in hippocampal volume and the tau measure of

Braak ROIs 1, 34 and 56 respectively among these cohorts. The volume of each side of the hippocampus was used as the predictor, and regional uptakes of Braak ROIs were used as the responses. Both sides of hippocampi have significant results with our federated Chow test. The p-values illustrate significant correlations between the changes of hippocampal volume and Tau deposition across all Braak ROIs and in all cohorts.

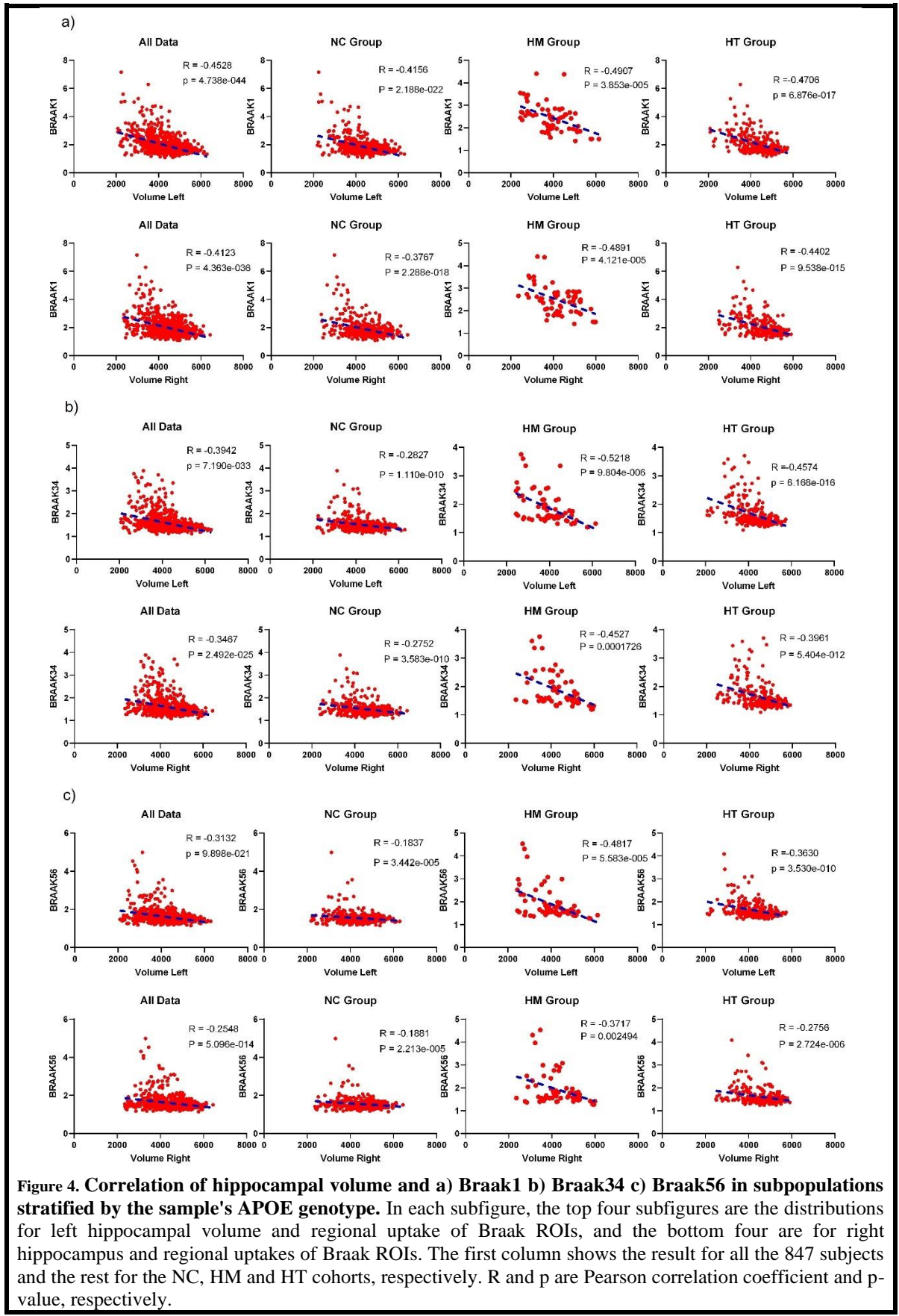

Figure 4. **Correlation of hippocampal volume and a) Braak1 b) Braak34 c) Braak56 in subpopulations stratified by the sample's APOE genotype.** In each subfigure, the top four subfigures are the distributions for left hippocampal volume and regional uptake of Braak ROIs, and the bottom four are for right hippocampus and regional uptakes of Braak ROIs. The first column shows the result for all the 847 subjects and the rest for the NC, HM and HT cohorts, respectively. R and p are Pearson correlation coefficient and p-value, respectively.

Pearson's correlation coefficients were also computed to evaluate the relationship between the hippocampal volume and regional uptakes within each subgroup, as illustrated in Figure 4. The top four subfigures are the distributions in left hippocampus, and the bottom four are for the right hippocampus. The first column shows the result for all the 847 samples, and the other three are for NC, HM and HT cohorts, respectively. R and p in each subfigure are the Pearson correlation coefficient and p-value. All the p-values are significant, and the cohort with HM genotype has the strongest negative correlation between volume and regional uptakes.

**3.2 Linking Hippocampal Surface Morphometry to Tau Deposition**

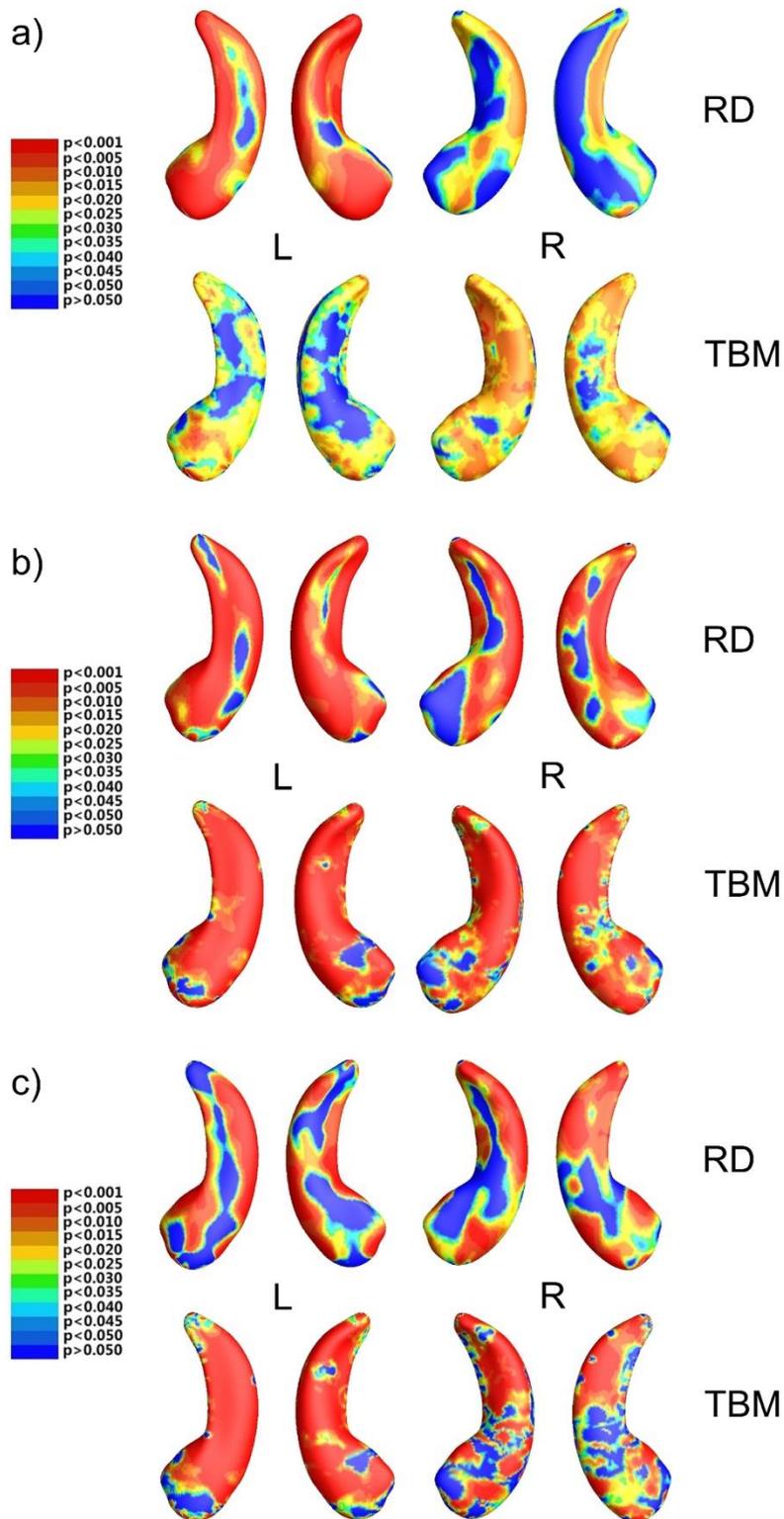

**Figure 5. The *p*-maps of our surface-based federated Chow test model results.** The warmer color regions have more significant *p*-values. a) Braak1; b) Braak34; c) Braak56. In each subfigure, the top two subfigures are the results for RD and the bottom two are for TBM. Braak1 results are less significantly related to tau measures under APOE-based stratification. The directionality information can be found in Figure 4.

We applied the proposed surface-based federated Chow test model to determine the regions where the atrophy focuses on the hippocampal surface. In this experiment, we computed two morphometry features, radial distance (RD) and tensor-based morphometry (TBM), in different cohorts. Morphometry features on each vertex is used as a predictor, and tau measures are used as a response. A Chow test *p*-value was calculated to evaluate the effect of tau deposition on the morphometry feature on each vertex across different *APOE* genotypes. The raw p-values were then adjusted for multiple comparisons according to the corresponding false discovery rates (FDR), and only those features with FDR < 0.05 were considered as functionally significant. As shown in **Table 2**, the average p-values of all combinations of RD/TBM and left/right hippocampi and Braak ROIs are significant.

Table 2 Critical p-values of all feature combinations with FDR < 0.05.

| Features | Braak1 | Braak34 | Braak56 |
|---|---|---|---|
| Left RD | 6.11e-8 | 1.49e-13 | 2.36e-14 |
| Left TBM | 1.59e-10 | 1.69e-13 | 1.85e-13 |
| Right RD | 4.20e-11 | 2.34e-14 | 4.44e-14 |
| Right TBM | 3.63e-10 | 1.07e-13 | 7.50e-14 |

Notice that although the p-values showed significance across all Braak ROIs, the *p*-maps illustrated a weaker correlation between Braak1 and tau measures under the APOE-based stratification condition. In addition, statistical evidence from previous studies revealed that significant cognitive deterioration first emerges in Braak34 stages [39,40]. Since biomarkers representing the initial deterioration stage would be more interesting, we focused our study on Braak34 and explore whether the correlated MRI features carry more significant statistical power for MCI conversion prediction in the following sections.

We further visualized the *p*-values on each hippocampal surface to identify the atrophy regions. Since the *p*-values were very significant, we normalized the *p*-values by dividing each *p*-value by the maximal *p*-value of each surface followed by a mapping of the values to the color map, as shown in **Figure 5**. Warmer colors represent lower p-values, and RD and TBM results were put on top and bottom sides in each subfigure, respectively. Specifically, the Braak34 results (b) illustrate an atrophy focus on the hippocampal subregions subiculum and cornu ammonis 1 (CA1 subfield), which is consistent with the previous studies [41–43].

**3.3 Cumulative Distribution Analysis**

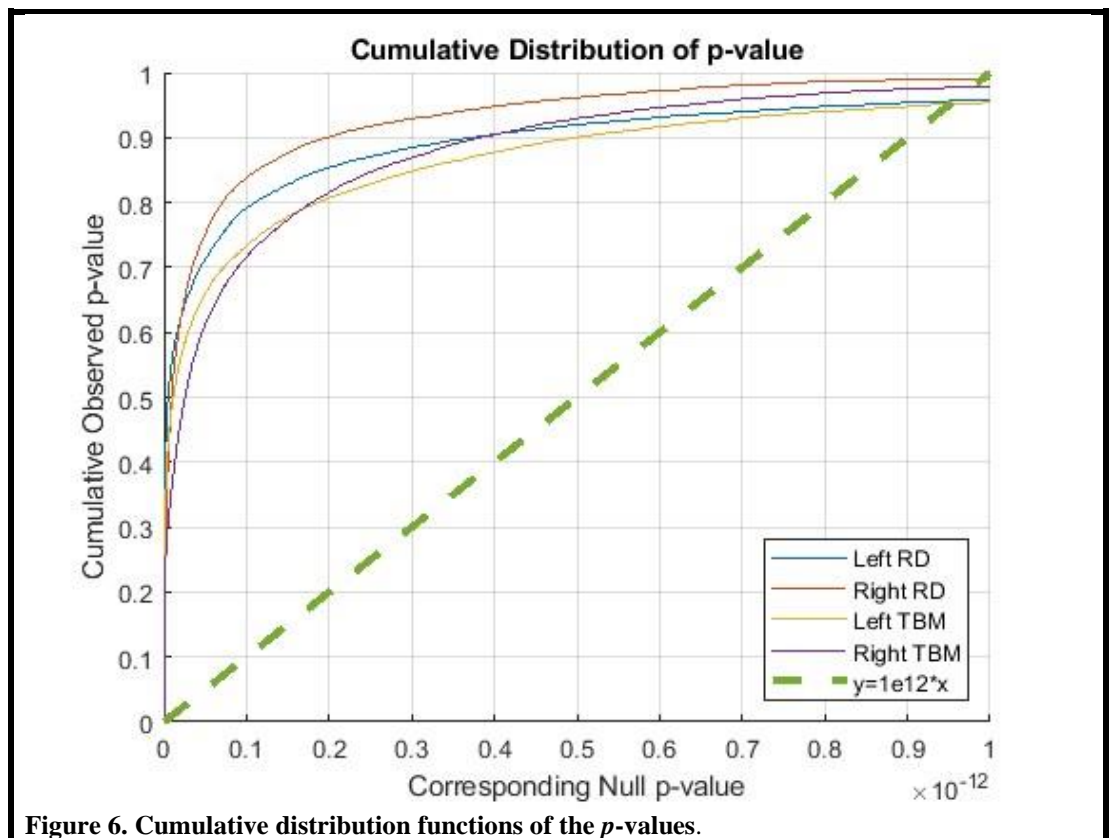

**Figure 6. Cumulative distribution functions of the *p*-values**.

To compare the effect sizes of different morphometry features, we created cumulative distribution function (CDF) plots of the previously computed *p*-values, as described in the our prior study [32]. For null distributions, the CDF of *p*-values is expected to fall approximately along the line (y = x). These empirical CDFs of *p*-values are the flip of the more common FDR PP plot, with steeper CDFs representing stronger effect sizes. Because both RD and TBM showed excellent effect size, it is hard to determine the better one with the null hypothesis ($y = x$). Instead, we used the $y = 10^{12}x$ line, where a significant effect is declared when the volume of suprathreshold statistics is more than $10^{12}$ times as that are expected under the null hypothesis. As shown in **Figure 6**, RD is slightly better than TBM.

**3.4 Predicting Clinical Decline in Participants with MCI**

In this experiment, we performed a survival analysis on another dataset [43] containing 110 MCI samples (56 converters and 54 non-converters) to evaluate the performance of our selected features (**Table 3**). Converters are defined as converting from MCI to AD within six years of diagnosis. Like Sect. 3.3, here we focused on the imaging biomarkers detected from the Braak34 correlation experiments. Additional efforts were made to ensure that none of the MCI subjects were included in the Braak34 correlation experiments.

We chose 50 RD and 50 TBM from the ROIs of each hippocampal surface and 50 features on the whole hippocampal surface to predict the conversion rates from MCI to AD, respectively. We adopt the previously determined best set of patch size of 10 by 10, while the fixed number of 50 RDs and 50 TBMs are chosen empirically for statistical tests with sample size of around 100. For comparison, we also performed the same experiment with the surface area and volume of the hippocampus. The hippocampal volume and surface area were calculated with the smoothed hippocampal structures after linearly registered to the MNI imaging space [23,31], and the sum of the bilateral hippocampal volume and the sum of the bilateral hippocampal surface area for each subject were used for this experiment.

For both random vertices in ROI and on the whole hippocampal surface, we calculated the average of the computed RD and TBM and used it as the scalar feature to fit the univariate COX model.

Table 3. Patient clinical information of 110 samples.

| Group | Sex (M/F) | Age | MMSE |
|---|---|---|---|
| MCI converter (n=56) | 36/20 | 75.1±6.8 | 27.0±1.6 |
| MCI non-converter (n=54) | 38/16 | 74.5±7.7 | 27.7±1.6 |

Values are mean ± standard deviation, where applicable.

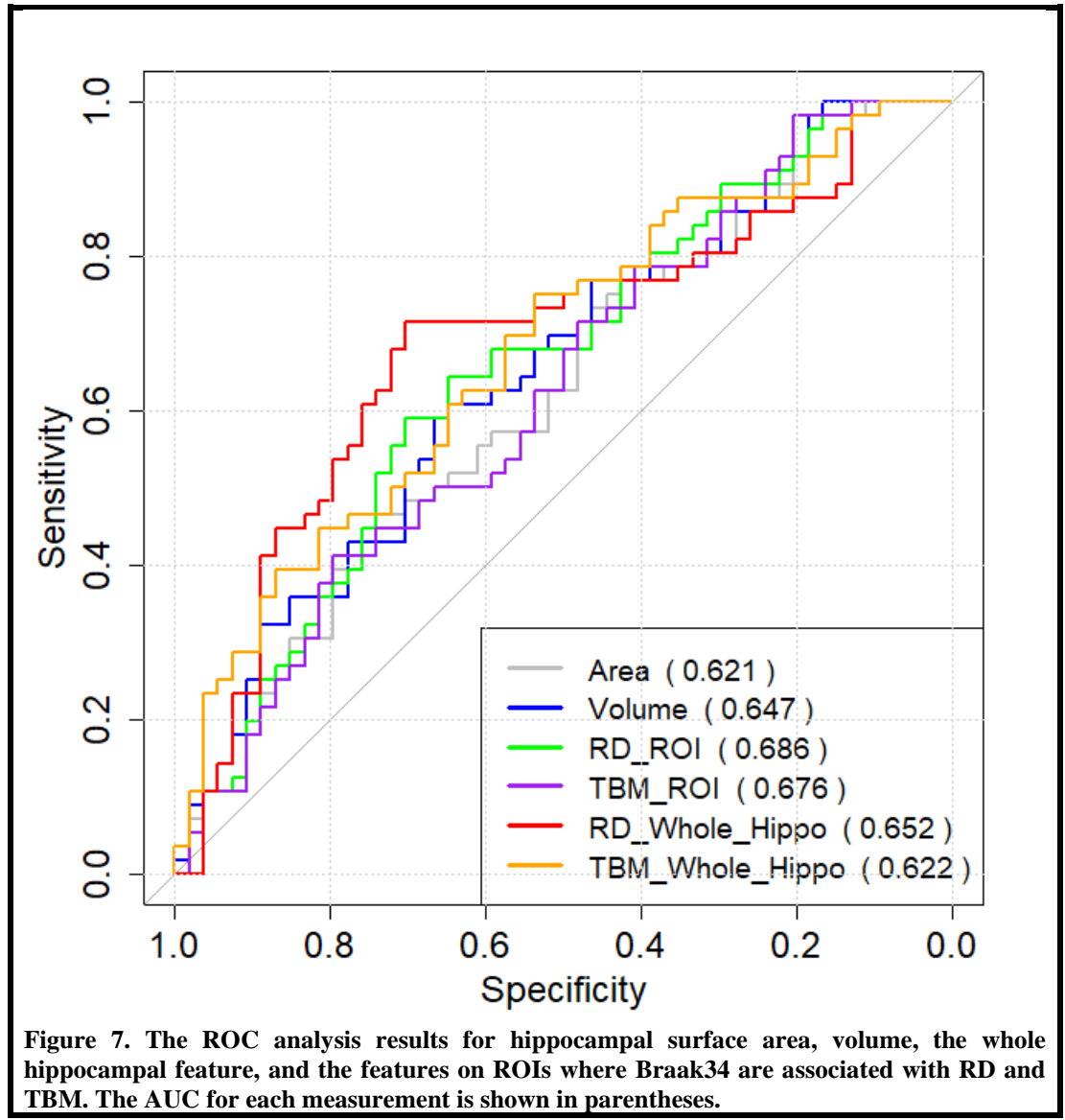

**Figure 7. The ROC analysis results for hippocampal surface area, volume, the whole hippocampal feature, and the features on ROIs where Braak34 are associated with RD and TBM. The AUC for each measurement is shown in parentheses.**

We then dichotomized the data by splitting the samples into high-value (HV) and low-value (LV) groups based on the optimal cutoffs of each measurement, followed by a comparative analysis of their effects on AD conversion. Specifically, the optimal cutoffs were determined using the maximum sum of sensitivity and specificity in ROC analysis [44](**Figure 7, Table 4**). As expected, AD may decrease the hippocampal volume and the other measurements. Next, we fit the Cox proportional hazard model [45] with both continuous values and the dichotomized version. The hazard ratios (HRs) and statistical significance (*p*-values) of both versions (HL for high/low dichotomization and COX for continuous) are shown in **Table 4**.

Moreover, we performed a Kaplan-Meier analysis to evaluate the difference of survival rates between the HV groups and LV groups. The features we used include the RD and TBM features on both whole hippocampal area and in ROIs, as well as surface area and volume (**Figure 8**). The p-values from log-rank test showed stronger significance with RD and TBM features on ROIs than other feature types.

**Table 4.** AUC for ROC Analysis, Optimal Cutoffs, and Estimated Hazards Ratios (HRs) for Conversion to AD in MCI Patients with High-value and Low-value Biomarkers Based on a Univariate Cox Model.

| Measurements | AUC (95% CI) | Cutoff | HL HR (95% CI) | HL *p*-value | COX HR (95% CI) | COX *p*-value |
|---|---|---|---|---|---|---|
| Area | 0.62 (0.52, 0.73) | 7447.9 | 2.1 (1.2, 3.7) | 5.84E-03 | >0.99 (>0.99, 1.0) | 1.04E-02 |
| Volume | 0.65 (0.54, 0.75) | 7712.9 | 2.4 (1.4, 4.1) | 1.35E-03 | >0.99 (>0.99, 1.0) | 1.21E-03 |
| RD_Whole_Hippo | 0.65 (0.55, 0.76) | 3.9 | 2.7 (1.6, 4.5) | 3.64E-04 | 2.1 (1.2, 3.6) | 6.46E-03 |
| TBM_Whole_Hippo | 0.62 (0.52, 0.73) | 3.9 | 2.1 (1.3, 3.6) | 5.47E-03 | 2.2 (1.3, 4.0) | 6.88E-03 |
| RD_ROI | 0.69 (0.58, 0.79) | 4.6 | 3.8 (2.1, 6.8) | 6.65E-06 | 2.1 (1.3, 3.2) | 1.11E-03 |
| TBM_ROI | 0.68 (0.58, 0.78) | 4.9 | 2.6 (1.4, 4.8) | 2.10E-03 | 2.2 (1.4, 3.3) | 2.68E-04 |

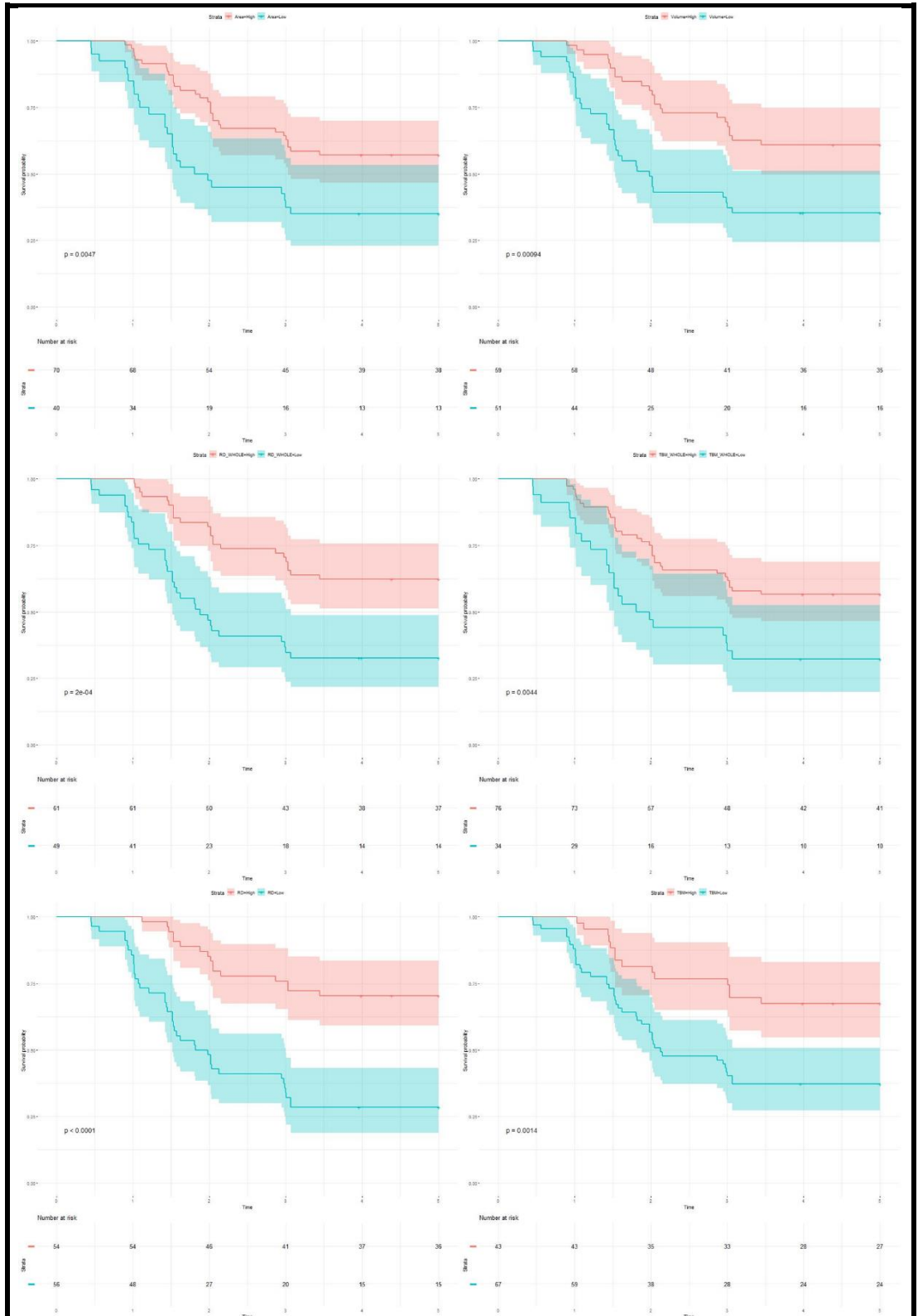

**Figure 8. The survival probability analysis for progression to AD in MCI patients based on hippocampal surface area, volume, the whole hippocampal features, and the features on ROIs related to Braak34.** The p-values are from the log-rank test. The red curve represents the high-value (HV) group for each measurement. And the blue one represents the low-value (LV) group.

### 3.5 Federated Learning Stability Analysis

Practically, different institutions might provide data with largely varying distributions. To demonstrate the stability of our model performance across different data distributions, we synthesized 1,000 samples and randomly assigned them to a set of independent hypothetical institutions varying in size, including one institution, three institutions, and five. As a result, our model yields unchanged sum of squared residuals (SSR) across linear regression models with different conditions (**Table 5**). The ground truth residual is shown in the first column, while the residuals under other conditions are shown in the rest columns. With the identical residuals, the resulting Chow test *F*-values remain the same, and the *p*-value will also be unchanged under different multi-site conditions. These results demonstrate the correctness and stability of our federated Chow test model.

**Table 5**. Stability analysis of federated linear model across different institutional settings.

|  | **Ground Truth** | **One-institution** | **Three-institution** | **Five-institution** |
|---|---|---|---|---|
| **SSR** | 3.96 | 3.96 | 3.96 | 3.96 |

## 4. DISCUSSION

This work generalizes our recent work [35] on the federated Chow test that identifies genetic and transcriptomic influences on brain sMRI measures. It proposes a surface-based federated Chow test model to detect the difference between tau deposition and hippocampal atrophy among cohorts with different genotypes. We performed various experiments with our model on the publicly available ADNI dataset and have three main contributions. The first contribution of this work is that it analyzes the difference of the associations of tau deposition and hippocampal atrophy among the cohorts with different genotypes. The framework may provide novel insights into the neural mechanisms about the impact of APOE and tau deposition on brain imaging and enrich the understanding of the relationship between hippocampal atrophy and AD pathology, and thus help in assessing disease burden, progression, and treatment effects. Secondly, by generalizing our prior work [35] to surfaces, the new model can extract the most AD-related regions on the hippocampal surface. The subiculum and cornu ammonis 1 (CA1 subfield) are identified as hippocampal subregions where atrophy is strongly associated with abnormal tau in the homozygote cohort. And the morphometry features on these selected regions show stronger correlations with AD progression. The proposed general method may be applied to analyze other brain imaging feature data. Finally, this model is built on a federated framework, which can investigate large datasets from different institutions without violating data privacy. A federated learning framework cannot only increase the statistical power but also can address the concern about the generalizability of out-of-sample data [10]. Although this study did not contribute methodologically to the federated framework, we decide to adopt the framework for practical biomedical considerations, because it is a constructive solution to the common issues of data privacy, scalability and cooperation among institutions.

**4.1 ROIs Related to Tau Deposition and Hippocampus**

Tau proteinopathies accelerate hippocampal atrophy leading to AD on MRI scans [21,41]. However, the influence of tau deposition on hippocampal morphology in the pathophysiological progression of AD is still not well understood. Some prior works [31,46,47] demonstrated that CA1 and the subiculum are the ROIs with the greatest abnormalities in the early stages of the AD pathophysiological process. Besides, the study of [41] reported a significant association between tau burden and atrophy in specific hippocampal ROIs (CA1 and the subiculum), and our recent work [13] also demonstrated the tau-related regions located at CA1 and the subiculum.

Our results are consistent with the prior studies noted above. Besides, our model can also demonstrate that the associations between tau deposition and hippocampal atrophy are most significant for the cohort with a homozygote genotype. The atrophy focuses on CA1 and the subiculum shown in **Figure 5**.

**4.2 Predictive Power of the Features on ROIs**

Our work selects ROIs by studying the trio relationship among AD induced brain structure, brain pathology, AD genotype. Our work is in line with recent work [48,49] that studied T-N relationship and discovered populations that may have imaging and clinical measures concordant with non-AD copathologies.

To verify the clinical values of these identified ROIs, we perform survival analysis (of conversion from MCI to AD) on the top features of the ROIs. Here, the univariate biomarker computed from our ROIs performs better than the traditional hippocampal volume, suggesting our ROIs' potential ability to study Alzheimer's disease as a univariate biomarker. A univariate biomarker may overcome inflated Type I error due to multivariate comparisons. For example, for randomized clinical trials (RCT), regulatory agencies, including the Food and Drug Administration (FDA), requires conventional univariate hypothesis testing and its associated statistical power analysis [50]. Consequently, our experimental results demonstrate the surface-based federated Chow test model's effectiveness in selecting the features related to AD progression. Our proposed model may enrich the understanding of the relationship among hippocampal atrophy, AD pathology, and AD genotype and eventually help assess disease burden, progression, and treatment effects.

**4.3 Economic considerations**

In addition to the difference of equipment availability between tau-PET and MRI scans, diagnosis cost is another issue of consideration, because the AD diagnosis tests on average are not budget friendly especially for elder population with limited income and/or without insurance. According to NewChoiceHealth.com (https://www.newchoicehealth.com), the average PET-scan cost in the United States is $5,750, while an MRI scan costs only $1,325 on average. Therefore, although MRI scans have moderately lower diagnosis accuracy, become a reasonable alternative for tau-PET in AD diagnosis. Our method not only improves the prediction power but also builds up a stronger association between MRI biomarkers and tau-

PET results, providing an opportunity for an improved MRI diagnosis accuracy and a remarkable reduction of patient expense on AD diagnosis.

**4.4 Limitations and Future Work**

Despite the promising results obtained by applying our federated Chow test model, there are three important caveats. First, this work only analyzes tau deposition. Beta-amyloid in the brain and blood-based biomarkers are also important hallmarks of AD pathology. Recent studies have associated these two biomarkers with brain structure atrophy [13,51]. In the future, we plan to conduct association analyses of hippocampal features and their relation to brain Beta-amyloid deposition and blood-based biomarkers. Second, hippocampus is not the only structure affected by AD pathology. With more datasets available, our framework will potentially make more contributions to other high-dimensional structures, such as the ventricles, and cortical surface metrics including gray matter thickness or volume [52,53]. Finally, besides APOE, our model can also analyze other genes and SNPs, which have been proved to be related to AD, like BIN1, CLU, and rs11136000 [6]. These future works will help shed new light on the relationship of component biological processes in AD.

**5. CONCLUSION**

This work proposes a novel surface-based federated Chow test model to investigate the effect of tau deposition and APOE on hippocampal morphometry in AD. Experimental results showed that our proposed model detect the difference in the association of tau deposition and hippocampal atrophy among the cohorts with different APOE genotypes and extract the most AD-related regions on the hippocampal surface, which focus on CA1 and the subiculum. And the top features on the selected ROIs show stronger predictive power in predicting AD progression. In the future, we will utilize our model to analyze more brain structures with different AD hallmarks, like brain amyloid deposition and blood-based biomarkers, and more AD-related genes and SNPs.

## 6. ACKNOWLEDGMENTS

Data collection and sharing for this project was funded by the Alzheimer's Disease Neuroimaging Initiative (ADNI) (National Institutes of Health Grant U01 AG024904) and DoD ADNI (Department of Defense award number W81XWH-12-2-0012). ADNI is funded by the National Institute on Aging, the National Institute of Biomedical Imaging and Bioengineering, and through generous contributions from the following: Alzheimer's Association; Alzheimer's Drug Discovery Foundation; BioClinica, Inc.; Biogen Idec Inc.; Bristol-Myers Squibb Company; Eisai Inc.; Elan Pharmaceuticals, Inc.; Eli Lilly and Company; F. Hoffmann-La Roche Ltd and its affiliated company Genentech, Inc.; GE Healthcare; Innogenetics, N.V.; IXICO Ltd.; Janssen Alzheimer Immunotherapy Research & Development, LLC.; Johnson & Johnson Pharmaceutical Research & Development LLC.; Medpace, Inc.; Merck & Co., Inc.; Meso Scale Diagnostics, LLC.; NeuroRx Research; Novartis Pharmaceuticals Corporation; Pfizer Inc.; Piramal Imaging; Servier; Synarc Inc.; and Takeda Pharmaceutical Company. The Canadian Institutes of Health Research is providing funds to support ADNI clinical sites in Canada. Private sector contributions are facilitated by the Foundation for the National Institutes of Health (www.fnih.org). The grantee organization is the Northern California Institute for Research and Education, and the study is coordinated by the Alzheimer's Disease Cooperative

Study at the University of California, San Diego. ADNI data are disseminated by the Laboratory for Neuro Imaging at the University of Southern California.

## 7. FUNDING

Algorithm development and image analysis for this study were partially supported by the ASU/Mayo Seed Grant Program, the National Institute on Aging (R21AG065942, U01AG068057, R01AG069453, RF1AG073424, and P30AG072980), the National Library of Medicine (R01LM013438), the National Institute of Biomedical Imaging and Bioengineering (R01EB025032), the National Eye Institute (R01EY032125), the National Institute of Dental & Craniofacial Research (R01DE030286), and the State of Arizona via the Arizona Alzheimer Consortium.

## 8. CONFLICT OF INTEREST

Yalin Wang and Richard J. Caselli are both members of the Editorial Board of this journal but were not involved in the peer-review process nor had access to any information regarding its peer-review. The remaining authors have no conflict of interest to report.

## 9. DATA AVAILABILITY

The data of this study are available from the corresponding author, Yalin Wang, upon reasonable request.

## 10. APPENDIX

---

**Algorithm 1. Federated Chow test Model.**

**Input:** Data pairs of the $I$ institutions, $(X_1, y_1), \ldots, (X_i, y_i), \ldots, (X_I, y_I)$ and the sample numbers of each group, $(N_1^{(1)}, N_1^{(2)}, N_1^{(3)}), \ldots, (N_i^{(1)}, N_i^{(2)}, N_i^{(3)}), \ldots, (N_I^{(1)}, N_I^{(2)}, N_I^{(3)})$

**Output:** $p$-value of Chow test

**Initialize:** $w^{(1)}, w^{(2)}, w^{(3)}, w^{(C)} = \mathbf{0}$

1: **for** $g = \{1,2,3,C\}$ **do**
2:     **while** *convergence and maximum number of iterations are not reached* **do**
3:         Each institution computes the gradient:
$$\nabla S_i^{(g)}(w^{(g)}) = [X_i^{(g)}]^T (X_i^{(g)} w^{(g)} - y_i^{g(g)}).$$
4:         Global center computes and sends global gradient to each institution:
$$\nabla S^{(g)}(w^{(g)}) = \sum_{i=1}^{I} \nabla S_i^{(g)}(w^{(g)}).$$
5:         Each institution updates the coefficient with the global gradient:
$$w^{(g)} \leftarrow w^{(g)} - \eta \nabla S^{(g)}(w^{(g)}).$$
6:     **end while**
7:     Each institution calculates the sum of squared residual: $S_i^{(g)}(w^{(g)}; X_i^{(g)}, y_i^{(g)})$.
8:     Global center gathers the global sum of squared residual: $S^{(g)} = \sum_{i=1}^{I} S_i^{(g)}$.
9:     Global center gathers the global sample numbers: $N^{(g)} = \sum_{i=1}^{I} N_i^{(g)}$.
10: **end for**
11: Global center calculates $F$ value with equation (1) and then computes and sends $p$-value to all institutions.

---